\documentclass[a4paper,letterpaper]{IEEEtran}
\usepackage{amsmath,amssymb,epsfig,enumerate,cite,footnote,authblk,color,hyperref}
\hypersetup{
    colorlinks,%
    citecolor=blue,%
    filecolor=blue,%
    linkcolor=blue,%
    urlcolor=blue
}
\definecolor{bluecolor}{rgb}{0,0.,1.}

\definecolor{redcolor}{rgb}{.7,0.,0.}

\newcommand{\pr}[1]{\left( #1\right)}
\newcommand{\prr}[1]{\left[ #1 \right]}
\newcommand{\es}[1]{\begin{equation}\begin{split}#1\end{split}\end{equation}}
\newcommand{\est}[1]{\begin{equation*}\begin{split}#1\end{split}\end{equation*}}

\newcommand{\R}{\mathbb{R}}

\newcommand{\V}{\mathcal{V}}

\newcommand{\rr}{\mathbf{r}}
\newcommand{\dd}{\textrm{d}}

\usepackage{fancyhdr}
\begin{document}
\title{\huge{Network connectivity in non-convex domains with reflections}}
\author[1,2]{Orestis Georgiou}
\author[1]{Mohammud Z. Bocus}
\author[2]{Mohammed R. Rahman}
\author[2]{Carl P. Dettmann}
\author[3]{Justin P. Coon}
\affil[1]{Toshiba Telecommunications Research Laboratory, 32 Queens Square, Bristol, BS1 4ND, UK}
\affil[2]{School of Mathematics, University of Bristol, University Walk, Bristol, BS8 1TW, UK}
\affil[3]{Department of Engineering Science, University of Oxford, Parks Road, OX1 3PJ, Oxford, UK}

\maketitle
\IEEEpeerreviewmaketitle

\pagestyle{plain}
\thispagestyle{fancy}

\begin{abstract}
Recent research has demonstrated the importance of boundary effects on the overall connection probability of wireless networks but has largely focused on convex deployment regions. 
We consider here a scenario of practical importance to wireless communications, in which one or more nodes are located outside the convex space where the remaining nodes reside. 
We call these `external nodes', and assume that they play some essential role in the macro network functionality e.g. a gateway to a dense self-contained mesh network cloud.
Conventional approaches with the underlying assumption of only line-of-sight (LOS) or direct connections between nodes, fail to provide the correct analysis for such a network setup. 
To this end we present a novel analytical framework that accommodates for the non-convexity of the domain and explicitly considers the effects of non-LOS nodes through reflections from the domain boundaries.
We obtain analytical expressions in 2D and 3D which are confirmed numerically for Rician channel fading statistics and discuss possible extensions and applications.
\end{abstract}


\section{Introduction \label{sec:intro}}

One fundamental requirement of wireless multi-hop relay or mesh networks is connectivity, i.e., that every node can communicate with every other node in a multihop fashion. 
Over recent years, a considerable amount of attention has been devoted to understanding the connectivity in such architectures (see \cite{akyildiz2005wireless,mao2012towards} and references therein).  
In the most simplistic and popular case, the network connectivity is analysed using the so called unit disk model where nodes can communicate only if they are within a specified distance from each other.
On the other hand, more practical models with channel fading statistics, irregular radiation patterns, and mobile nodes have been also studied \cite{Bettstetter2005, Miorandi2008,georgiou2013connectivity}. 
Finally, the impact of boundaries in dense networks has also been investigated in \cite{Coon2012}, where it is shown that in the dense regime, the geometry and shape of the network deployment region plays a dominant role in shaping the network topology.

\begin{figure}[t]
\centering
\includegraphics[scale=0.19]{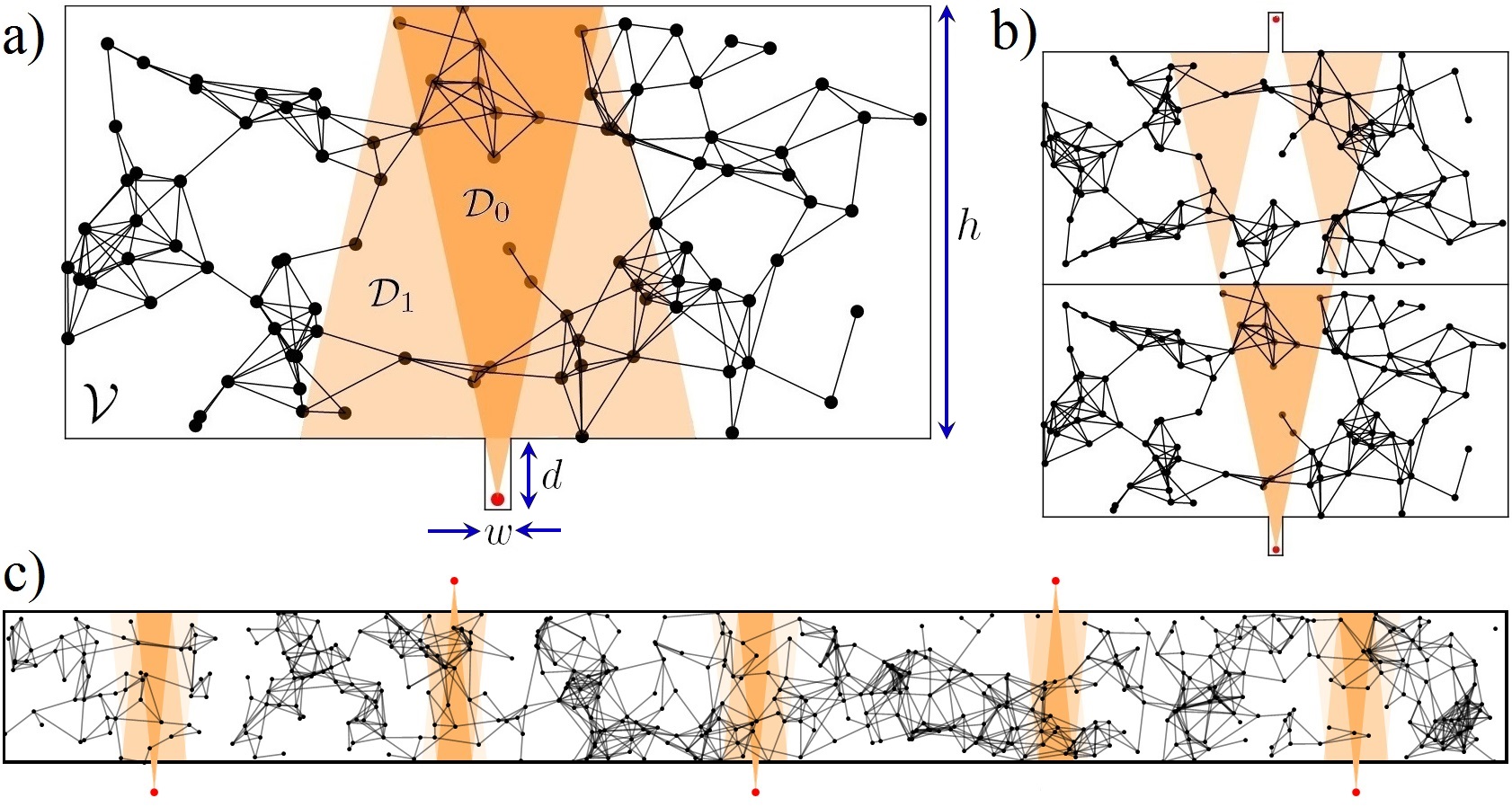}
\caption{a) Illustration of the basic system setup showing a rectangular domain $\V$ with $N=100$ interior nodes (black dots) and an external node (red dot) placed at the bottom of a small hole. The figure also illustrates the line of sight region $\mathcal{D}_0$ and the first reflection region $\mathcal{D}_1$. b) Unfolded setup (reflected along the top parallel wall) as to illustrate an easy way of visualising the different connection regions $\mathcal{D}_c$. c) WCN example with five external nodes connecting to the internal mesh network of $N=500$.}
\label{fig:setup}
\end{figure}

Although the connectivity of random networks has been extensively studied, most existing work has considered only convex geometries.
In other words, there is always an assumption that a connection between two nodes is only possible if a line of sight (LOS) exists.
In this letter, we study a scenario where such assumptions need not hold and to the best of our knowledge, together with \cite{georgiou2013network}, constitutes the first analytic approach to network connectivity in non-convex domains. 
Significantly, we investigate connectivity in a network where one or more nodes reside outside the main system e.g. in small holes situated in the boundaries of the confining geometry as illustrated on the left of Fig. \ref{fig:setup}. 
For instance, such a scenario may model wireless sensors placed in a ditch, trench, or drain of some underground road or train tunnel tasked with reporting measurements to a mesh network.
Alternatively, one may consider wireless access points placed in hidden-from-view or obstructed indoor locations, commonly found in complex industrial environments \cite{savazzi2013wireless}.
Such setups are the focus of the novel wireless communication paradigm referred to as a Wireless Cloud Network (WCN) where messages are flooded from sensors/actuators through a self-contained dense mesh network (the cloud) to the backhaul via so called cloud access nodes i.e. the external nodes \cite{savazzi2013wireless}. 
Hence, external nodes may have some essential functionality or network role and should therefore be well connected to the interior mesh.

Considering that in practical situations, wireless connectivity in confined geometries may involve dominant reflected signals from boundaries, we present here a means of modelling these reflected rays importing ideas from mathematical billiards \cite{dettmann2009survival} and using them in conjunction with Rician channel fading statistics. 
We obtain expressions for the connectivity of external nodes to the internal mesh and confirm them numerically, thus enabling the analysis of network connectivity in non-convex geometries with reflection effects.
The current analysis lends itself to WCN applications such as smart metering and industrial monitoring where a dense mesh topology is often randomly defined in non-convex deployment regions.

\section{Network Definitions and System Model \label{sec:model}}

We begin by considering a rectangular domain $\V\subset\R^2$ of area $V$ and height $h$ which also contains a small hole of width $w\ll1$ and depth $d$ on one of it's walls (see Fig. \ref{fig:setup}).
There are a total of $N$ nodes randomly and uniformly scattered in $\V$ with density $\rho=N/V$ and position coordinates given by $\rr_i\in\V$ for $i=1,2,\ldots N$.
These nodes represent wireless devices which can connect to form a network of communication links according to some appropriate connection model.
A popular model for example has been the so called ``unit disk" model where two nodes connect  if they are within a certain distance from each other.
Recently, the ``random connection" model has been adopted where more realistic effects such as path loss, channel randomness, multi path fading, and antenna characteristics (e.g. power, gain, number of antennas etc.) can be incorporated into the pair connection function $H_{ij}$ in a consistent way.
Namely, $H_{ij}$ can be viewed as the complement of the information outage probability \cite{Proakis2000}
$
H_{ij}=1-P_{out} = P\left(\log_2(1+\text{SNR}\times |h_{ij}|^2) > R_0 \right) 
$
where $|h_{ij}|^2$ is the channel gain between two nodes, $R_0$ is the minimum information rate and the SNR is the long time average signal-to-noise ratio at the receiver. From the Friis transmission formula we have that
$
\text{SNR} \propto G_T G_R r^{-\eta} \alpha^{c} ,
$
where $G_T$ and $G_R$ are the antenna gains at the transmit and receive nodes respectively, $r_{ij}$ is the dimensionless distance (relative to the signal wavelength) between the nodes in question, $\alpha\in[0,1]$ is the dimensionless factor representing the amount of signal attenuation upon collision/reflection with the boundaries \cite{Zhao2001}, $c$ is the number of reflections in the signal trajectory and $\eta$ is the environment dependent path loss exponent.  
Typical values of $\eta$ are $2$ under the free space path loss assumption and $\eta>2$ for dense or crowded environments.
For simplicity, we will assume isotropic antennas such that $G_T=G_R=1$.

Statistical models for the random variable $h_{ij}$ vary considerably according to the system's underlying assumptions from simple Rayleigh fading to heavy tailed distributions such as log-normal fading channels.
Here we are predominantly interested in understanding line of sight (LOS) and reflection effects and hence adopt a Rician fading model \cite{Proakis2000} where a strong signal can be observed at the receiver along with other weaker signals. 
The stronger signal typically is the one that traverses the shortest distance from transmitter to receiver (and \textit{vice versa}), and also undergoes fewer reflections from attenuating walls.
The channel power density of a Rician fading channel is given by 
$
f_X(x) = \frac{(K+1)}{\omega} e^{-(K+\frac{(K+1)x}{\omega})}I_0 \! (\sqrt{4K(K+1)x/\omega} )
,$
where $K$ is the Rice factor, generally taken to be greater than unity (with typical values around 4 or 5 being reasonable for this work), $\omega$ is a channel dependent parameter and $I_0$ is the modified Bessel function of the first kind.  The cumulative distribution function is therefore
$
F_X(x) = 1 - Q_1(\sqrt{2K},\sqrt{2(K+1)x/\omega} ),
$
where $Q_1(a,b)$ is the Marcum $Q$-function.  It follows that the connection probability for a pair of nodes is 
\es{
H_{ij}^{(c)} = Q_1\left(\sqrt{2K},\sqrt{2(K+1) \beta r_{ij}^{\eta} \alpha^{-c}} \right),
\label{H}}
where $\beta =1/r_0^\eta$ depends on transceiver parameters such as frequency and transmit power, and $r_0$ is interpreted as the typical distance at which a good communication link can be established.
Note that we use superscript $(c)$ to indicate the number of reflections the shortest path signal has experienced.

Finally, we detail the main focus of this paper which is an external node $\rr_k$ which is situated at the bottom of the small hole as illustrated in Fig. \ref{fig:setup}.
This node clearly has a restricted LOS (a keyhole-like) view of the rest of the network, which is about $\phi=2\arctan \frac{w}{2d}$ radians. 
We denote the LOS region as $\mathcal{D}_0$ such that $|\mathcal{D}_0|\approx \phi h^2/2$ when $w\ll d\ll 1$.
Such geometric characteristics break the convexity of the domain $\V$ and beg further investigation in understanding how they affect the connectivity of the global network.
In the following, we will abstract the specific functionality of external nodes in order to study the geometrical impact of non-LOS and reflection effects to the local and global network connectivity.

\section{Keyhole and Reflection Effects \label{sec:full}}

Full connectivity in a network is a global observable and is achieved if at any instance, every node can communicate with every other node in a multihop fashion. 
While a very strong measure of connectivity, its compatibility with delay and/or disruption tolerant networking can increase wireless network reliability and prevent disruptions due to factors like radio range, node sparsity, energy resources, attack and noise \cite{fall2008dtn}.
Recently, it was shown in \cite{Coon2012} that the probability of achieving full connectivity in random networks denoted here by $P_{fc}$ is dominated by ``hard to connect to" regions e.g. when nodes are situated near sharp corners of the domain geometry. 
This general conclusion remains valid in the present setting of a hard to connect to exterior node $k$ placed near a small hole in the domain's boundary.
We therefore expect that the above defined network is fully connected if the $N$ random nodes in $\V$ form a connected cluster, \textit{and} at least one of them is directly connected with the exterior node at $\rr_k$
\est{
P_{fc}&=P_{fc}^{(\V)} \text{Pr}(\text{at least one link with node } k)\\
&=P_{fc}^{(\V)} \prr{1- \text{Pr}(\text{no links with node } k)}\\
&= P_{fc}^{(\V)} \Big( 1-\langle \prod_{i=1}^{N}(1-H_{ki}) \rangle \Big)  \approx P_{fc}^{(\V)} \Big( 1- e^{- N \langle H_{ki} \rangle } \Big)
,}
where the angled brackets $\langle O \rangle = \frac{1}{V^N} \int_{\V^N} O \, \dd \rr_1 \ldots \dd \rr_N$ indicate the unconditional expectation probability of event $O$ obtained through a spatial average over all possible network realizations.
The approximation symbol used in the last line indicates that we have have used the
fact that $(1-x)^N \approx e^{-Nx}$ for $N \gg 1$ and $x \ll 1$.
We have also assumed that the connection probability of links $(k,i)$ and $(k,j)$ are sufficiently independent such that the product turns into a power.
For the remainder of the paper we will focus on the calculation of $\langle H_{ki} \rangle= \frac{1}{V} \int_{\V} H_{ki} \dd \rr_i$ since $P_{fc}^{(\V)}$ for $ \V$ convex has been discussed in some detail in \cite{Coon2012}.

From the definition of $\eqref{H}$ we can express the probability that node $k$ connects with some randomly selected node as a sum of direct and reflected contributions up to some maximum number of reflections $C$ as
\es{
\langle H_{ki} \rangle = \frac{1}{V} \sum_{c=0}^{C} \int_{\mathcal{D}_c} H_{ki}^{(c)} \dd \rr_i
\label{HK}.}
The respective integration regions $\mathcal{D}_0$ and $\mathcal{D}_1$ are shown in Fig. \ref{fig:setup}.
It can be seen that for parallel wall reflections $|\mathcal{D}_c| \approx 2| \mathcal{D}_0 |$ for $c\geq 1$ (subject to the width of $\V$ being large enough) and therefore we expect a trade-off between a greater (about double) connection region and the attenuated reflected signal.
We will confirm our results later by comparing $ \langle H_{ki} \rangle$ with numerical simulations of the mean degree of the external node (see Fig. \ref{fig:Hie}) given by $\mu_k = N\langle H_{ki} \rangle$.

\subsection{Line of sight and $\mathcal{D}_0$}

We first turn to the LOS signal which can only reach nodes in region $\mathcal{D}_0$. 
These nodes are potential candidates for establishing a direct unattenuated link with the external node $k$.
The expected probability of such an event is therefore given by $\frac{1}{V} \int_{\mathcal{D}_0} H_{ki}^{(0)} \dd \rr_i$.
Analytical integration of the Marcum $Q$ pair connectedness function in the form of \eqref{H} is impossible, and therefore we resort to approximations presented in \cite{Bocus2013}
\es{\label{Qapprox}
Q_1(a,b) \approx \exp\pr{ -  e^{\nu(a)}   b^{\mu(a)} }
,}
where $\mu(a)= 2.174 - 0.592a + 0.593a^2 - 0.092a^3 + 0.005a^4$
and $\nu(a)= -0.840 + 0.327a - 0.740a^2 + 0.083a^3 - 0.004a^4$.
A second difficulty arises from the triangular shape of $\mathcal{D}_0$. 
To make progress we approximate the triangular shape of $\mathcal{D}_0$, for $w\ll d \ll 1$ (i.e. when $\phi\ll 1$) with a circular sector of similar opening $\phi$ and radius height $h$.
Adopting such an approximation we are now faced with a radially symmetric integrand which can be calculated in closed form to give
\es{
 \int_{\mathcal{D}_0}\!\! H_{ki}^{(0)} \dd \rr_i & = \phi \int_{0}^{h} \!\! r e^{-\lambda_0 r^{\kappa}} \dd r = \frac{\phi}{\kappa \lambda_0^{2/\kappa}} \gamma\left(\frac{2}{\kappa},\lambda_0 h^{\kappa} \right),
\label{eq:H_01_full}
}
where $\phi$ is the angle available from $\rr_k$ to other nodes, $\lambda_c = e^{\nu(a)} (2(K+1)\beta\alpha^{-c})^{\mu(a)/2}$, $a=\sqrt{2K}$, $\kappa=\mu\eta/2$ and $\gamma(.,.)$ is the lower incomplete gamma function.
Notice that the sector approximation implies that \eqref{eq:H_01_full} underestimates the true probability with an error of only $\mathcal{O}(\phi^3)$.

\subsection{Reflection effects \label{sec:reflections}}
\begin{figure}[t]
\centering
\includegraphics[scale=0.32]{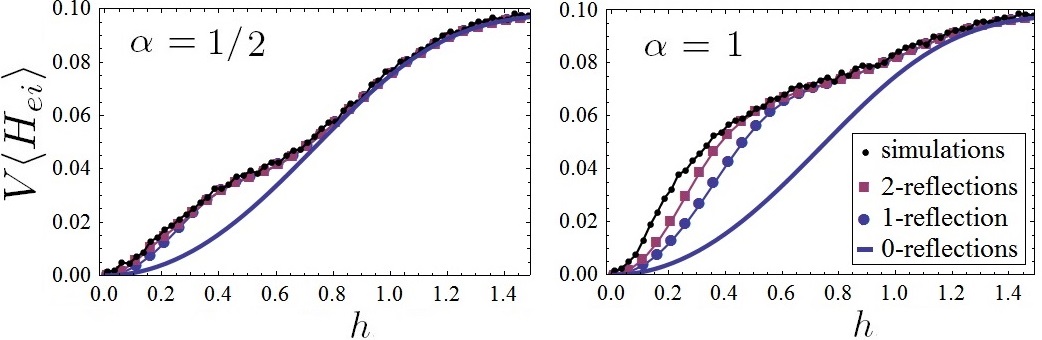}
\caption{The un-normalized probability that the external node $\rr_k$ connects with some random node $V\langle H_{ki} \rangle$ given in \eqref{HK} as a function of the domain height $h$.
The solid blue curve is the result of \eqref{eq:H_01_full} ignoring any reflected signals. 
The blue and purple points involve the numerical integration of \eqref{eq:H_0c_full} for $C=1,2$, i.e. accounting for 1 and 2 reflections respectively.
The black dotted curve is obtained through numerical simulations which calculate the mean degree $\mu_k$ divided by the simulated node density $\rho=N/V$.
Parameters used: $\beta=1$, $K=4$, $\eta=2$, and $\phi=\pi/16 \approx 0.196$.}
\label{fig:Hie}
\end{figure}

After the first reflection from the parallel wall opposite the external node, nodes in $\mathcal{D}_1$ can connect with $\rr_k$ even though the received signal is attenuated by $\alpha$ as in \eqref{H}.
A nice way of visualizing the integration region is by ``unfolding" the domain along the horizontal top boundary of $\V$ as illustrated in Fig. \ref{fig:setup}.b).
This unfolding trick is commonly used in mathematical billiards \cite{dettmann2009survival} and is essentially just a change of variables transformation which considerably simplifies our analysis (particularly for $c\geq1$).
Notice that $\mathcal{D}_1$ in the unfolded picture consists of two inverted isosceles triangles.
Therefore, when invoking the sector approximation as above we have to exclude nodes already considered in $\mathcal{D}_0$.
The upper limit of the radial integral hence is constant at $2h$, however the lower one is some function of the polar angle $\theta$ describing the diagonal straight line through $2h$ and gradient $-\cot\phi/2$ giving
\es{
\int_{\mathcal{D}_1} &H_{ki}^{(1)}  \dd \rr_i = 2 \int_{\frac{\pi-\phi}{2}}^{\frac{\pi}{2}} \int_{r_{1}}^{2h} r H_{ki}^{(1)} \dd r \dd \theta \\
&= \frac{2}{\kappa \lambda_1^{2/\kappa}} \bigg[ \frac{\phi}{2}  \gamma\left(\frac{2}{\kappa},\lambda_1 (2h)^{\kappa} \right) 
- \int_{\frac{\pi-\phi}{2}}^{\frac{\pi}{2}} \! \gamma\left(\frac{2}{\kappa},\lambda_1 r_{1}^{\kappa} \right)  \dd \theta \bigg] \\
&< \frac{\phi}{\kappa \lambda_1^{2/\kappa}} \bigg[  \gamma\left(\frac{2}{\kappa},\lambda_1 (2h)^{\kappa} \right)- \gamma\left(\frac{2}{\kappa},\lambda_1 h^{\kappa} \right) \bigg]
\label{eq:H_02_full}
,}
where $r_{1}(\theta) = 2h\sin(\phi/2) \sec(\theta-\phi/2)$, and the inequality follows by  setting $r_{1}=h$.
Generalizing \eqref{eq:H_02_full} to $c\geq 1$ we get
\es{
\int_{\mathcal{D}_c} H_{ki}^{(c)} \dd \rr_i & = 2 \int_{\frac{\pi-\phi}{2}}^{\frac{\pi}{2}-\phi_c} \int_{r_{c}}^{(c+1)h} r H_{ki}^{(c)} \dd r \dd \theta \\
&=\frac{2}{\kappa \lambda_c^{2/\kappa}} \bigg[ \frac{\phi-2\phi_c}{2}  \gamma\left(\frac{2}{\kappa},\lambda_c ((c+1)h)^{\kappa} \right) \\
&\qquad - \int_{\frac{\pi-\phi}{2}}^{\frac{\pi}{2}-\phi_c} \! \gamma\left(\frac{2}{\kappa},\lambda_c r_{c}^{\kappa} \right)  \dd \theta \bigg] 
\label{eq:H_0c_full}
,}
where $r_{c}= 2ch\sin(\phi/2) \sec(\theta-\phi/2)$, and $\tan\phi_c = \frac{c-1}{c+1}\tan\frac{\phi}{2}$. 
Notice that all reflected contributions to \eqref{HK} (i.e. $c\geq 1$) are unimodal functions of $h$.
We therefore briefly investigate the maximum by looking at the upper bound of \eqref{eq:H_0c_full} obtained by setting $r_{c}= ch$. 
Differentiating with respect to $h$ and solving we obtain that $h_c^{\text{max}} = \sqrt[\kappa]{\frac{\ln 4 }{\lambda_c (2^\kappa -1)}}$.

Fig. \ref{fig:Hie} shows a plot of  $V \langle H_{ki} \rangle $ in \eqref{HK} as a function of the domain height $h$, for $C=0,1$, and $2$, using different values of the attenuation factor $\alpha$.
The scale is set by the value of $\beta=1/ r_0^\eta$, such that when $h\gg r_0$ reflection effects are minuscule regardless of the attenuation $\alpha$ as reflected signals decay exponentially with $h$.
Reflection effects can however have a considerable effect to the connectivity of the external node for narrow domains $\V$ (i.e. $h\ll r_0$) especially when attenuation is low (i.e. $\alpha \approx 1$). 
Fig. \ref{fig:Hie} also includes results from computer simulations which validate the above analysis. 
Significantly, including just a few reflections in \eqref{HK} seems to be enough to achieve good agreement with simulation results.

\section{Extensions and Generalizations \label{sec:extra}}

\subsubsection{3-D keyhole}
We present here the generalization of the above results to a 3D domain $\V\subset \R^3$, by considering a rectangular cuboid with a small circular hole of opening diameter $\hat{w}$ and depth $d$ such that the maximum polar angle of the LOS view of the network (analogous to $\phi/2$ in the 2D case) is now $\psi= \arctan \frac{\hat{w}}{2 d}$.
Therefore the total solid angle available to the external node is $\varphi=2\pi (1- \cos\psi)$.
Assuming that $\hat{w}\ll d \ll 1$ and that the Marcum-Q function is approximated as in \eqref{Qapprox} we can calculate in polar coordinates the line of sight approximate contribution
\est{
\int_{\mathcal{D}_{0}} \!\! H_{ki}^{(0)} \dd \rr_i & = 2\pi \! \int_{0}^{\psi} \!\! \int_{0}^{h} \!\! r^2 \sin \theta e^{-\lambda_0 r^\kappa} \dd r \dd \theta 
=  \frac{\varphi \gamma\left( \! \frac{3}{\kappa},\lambda_0 h^{\kappa} \! \right)}{\kappa \lambda_0^{3/\kappa}} 
.}
Similarly, we may calculate the 
reflection contributions
\es{
\int_{\mathcal{D}_c} \!\! H_{ki}^{(c)} \dd \rr_i &= 2\pi \int_{\psi_c}^{\psi} \int_{r_{c}}^{(c+1)h} \!\!\! r \sin\theta  H_{ki}^{(c)} \dd r \dd \theta 
,}
where $r_{c}= 2ch\cos(\psi) \csc(\theta-\psi)$, and $\tan\psi_c = \frac{c-1}{c+1}\tan\psi$. 
A similar behaviour as that shown in Fig. \ref{fig:Hie} is observed with $h_c^{\text{max}} = \sqrt[\kappa]{\frac{\ln 8 }{\lambda_c (2^\kappa -1)}}$.
Notice that the above 3D calculation holds for any shaped hole and not just a circular one, as long as the LOS total solid angle is equal to $\varphi$. This can easily be confirmed by considering a square hole.
Interestingly, and unlike the 2D case, $|\mathcal{D}_c| = 6 c |\mathcal{D}_0|= 2 c h^3 \varphi$ implying that reflection effects are more significant in 3D due to an increase in the available reflected connection regions.

\subsubsection{Multiple holes}
The above simple system setup can be generalized to accommodate for multiple holes with multiple external nodes.
The results follow from previous expressions:
\es{
P_{fc} = P_{fc}^{\V} \prod_{k} \pr{1- e^{-\rho \langle H_{ki} \rangle}}
,}
where the product runs over all external nodes $k$ and we have assumed that the connection probabilities of external nodes are sufficiently independent.
This is sensible if the LOS regions $\mathcal{D}_0$ of the external nodes have no overlaps.

\begin{figure}[t]
\centering
\includegraphics[scale=0.3]{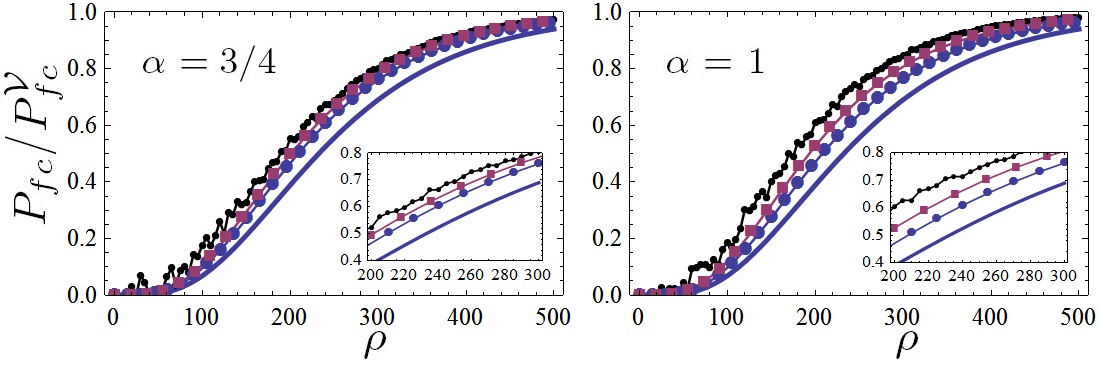}
\caption{The probability that five external nodes are all connected with the internal mesh network (as in Fig. \ref{fig:setup}.c) with $h=0.3$ and length $L=5$ such that $V=1.5$)) as a function of the node density $\rho=N/V$. The insets highlight the region $\rho\in[200,300]$. Curves and parameters are as in Fig. \ref{fig:Hie}.}
\label{fig:multi}
\end{figure}

Indeed, this setup approaches that of a WCN described in Sec. \ref{sec:intro} (see also Fig. \ref{fig:setup}.c)).
In Fig. \ref{fig:multi} we plot $P_{fc}/P_{fc}^{\V}= \pr{1- e^{-\rho \langle H_{ki} \rangle}}^5$ as a function of the density $\rho=N/V$ for the setup shown in Fig. \ref{fig:setup}c) and parameters as in Fig. \ref{fig:Hie}. 
Clearly this tends to $1$ as $\rho\to\infty$.
The exponent $\langle H_{ki} \rangle$ is calculated using \eqref{HK} for $C=0,1,2$. 
The numerical simulations shown as black points are the expected probability that all five external nodes are connected with the interior mesh network of density $\rho$. 
Comparing the 0-reflections curve to simulations, it is clear that incorporating reflection effects into \eqref{HK} can have a significant effect (up to $40\%$) in the current setup.

\section{Conclusions and Discussion \label{sec:conc}}
 
In this contribution, we imported ideas from mathematical billiards in order to study reflection effects in non-convex wireless network deployment regions; a setup which may find application in many ad hoc use cases including so called wireless cloud networks (WCN) \cite{savazzi2013wireless}.
We study an external node with restricted LOS view which communicates with an internal wireless mesh and derive general expressions in both 2D and 3D geometries for local and global connectivity thus enabling the analysis of network connectivity in non-convex domains.
The statistical framework that we adopt can account for a broad range of small and large-scale fading models.  
Here, we consider a Rician model, which accounts for a single dominant component of the signal. 
We do not explicitly account for antenna directivity, however other models (e.g. \cite{durgin2002new}) could instead be applied although would require a more thorough investigation (perhaps using techniques disclosed in \cite{georgiou2013network})
and is not within the scope of the present submission.
Another possibility for further work could entail understanding the impact effect of reflections from curved boundaries.

\section*{Acknowledgments}
The authors would like to thank the EPSRC (grant EP/H500316/1), the FP7 DIWINE project (Grant Agreement CNET-ICT-318177), and the directors of the Toshiba Telecommunications Research Laboratory for their support.



\bibliographystyle{ieeetr}
\bibliography{mybib}

\end{document}